# Non-destructive and Rapid Evaluation of CVD Graphene by Dark Field Optical Microscopy

X. H. Kong, H. X. Ji, R. D. Piner, H. F. Li, C. W. Magnuson, C. Tan, A. Ismach, H. Chou, R. S. Ruoff [a]

*Department of Mechanical Engineering and the Materials Science and Engineering Program, The University of Texas at Austin, 1 University Station C2200, Austin, Texas 78712, United States*

[a] Author to whom correspondence should be addressed. E-mail: r.ruoff@mail.utexas.edu

**ABSTRACT**: Non-destructive and rapid evaluation of graphene directly on the growth substrate (Cu foils) by dark field (DF) optical microscopy is demonstrated. Without any additional treatment, graphene on Cu foils with various coverages can be quickly identified by DF imaging immediately after chemical vapor deposition growth with contrast comparable to scanning electron microscopy. The improved contrast of DF imaging compared to bright field optical imaging was found to be due to Rayleigh scattering of light by the copper steps beneath graphene. Indeed, graphene adlayers are readily distinguished, due to the different height of copper steps beneath graphene regions of different thickness.

**Introduction**

The discovery of large area monolayer growth on Cu foils has provided a route to the synthesis of very large area graphene.[1-5] Large-area graphene is typically a polycrystalline layer that contains various types of defects and sometimes adlayers. Non-destructive and rapid identification of defects, grain boundaries, and adlayers on the growth substrate is useful for understanding graphene growth and for assessing the quality of graphene samples of large lateral size, and thus also for metrology and the industrial scale production of graphene. Typically, scanning electron microscopy (SEM) and Raman spectroscopy (including Raman mapping) have been used to map the structure of graphene films. However, SEM imaging typically introduces an amorphous carbon coating on the exposed region,[6] and micro-Raman mapping is so time consuming that it is limited to small regions of a graphene film.[7]

Optical microscopy has been explored for imaging graphene and the image 'quality' for discerning critical features depends on the optical properties of the substrate and the wavelength used.[8, 9] Typically, graphene has been transferred onto a Si substrate with a ~280 nm thick thermal oxide layer and this oxide thickness favors identification by white light illumination.[8, 9] Deposition and alignment of a thin liquid-crystal layer on graphene has been used to detect graphene grains and grain boundaries by optical birefringence.[10] Thermal annealing in air[11] or irradiation with ultraviolet (UV) light under moisture-rich ambient conditions[12] yielded optical images of graphene grain boundaries as well as separated domains (sometimes also referred to as islands) on Cu foils owing to the increase of interference color contrast between copper oxide and Cu[13]. All of these optical microscopy methods used bright field (BF) imaging, and the brightness and color imaging contrast arise from different degrees of light absorption. Of course, the UV-assisted oxidation damages the graphene and is thus a destructive method[12]. A fast, inexpensive, and contamination-free mapping of graphene on the growth substrate is needed. Lewis *et al.*[14] have used dark field (DF) optical microscopy to detect graphene. Our work covers a variety of additional aspects.

We report here that freshly grown graphene on Cu foils with various coverages from sub-monolayer to fully-covered can be rapidly identified by dark field (DF) optical imaging without any additional treatment. The DF imaging of graphene on Cu presents a contrast comparable to that of SEM imaging with sub-micrometer resolution owing to Rayleigh scattering of light by the copper steps beneath graphene. In addition, different height copper steps beneath graphene can be identified by DF imaging and we have found that, due to this, adlayers are also distinguishable with DF imaging.

**Results and Discussion**

Graphene was synthesized on Cu foils (25-µm thick, 99.8% Alfa Aesar no. 13382) by low pressure chemical vapor deposition (CVD) technique as previously reported.[1, 3] Growth was done in a hot wall tube furnace at 1030 °C with methane ($CH_4$; 99.999%, Air Gas, Inc.) and hydrogen ($H_2$; 99.999%, Air Gas, Inc.) as precursors. For the convenience of locating same graphene domains to compare DF optical imaging with other imaging methods, partial coverage as opposed to full coverage graphene on Cu was the primary focus of our study, but graphene on Cu at full coverage was studied as well. Optical microscope images were obtained with a Zeiss Axiovert 100A light microscope equipped with a dark-field condenser and a tungsten lamp. SEM images were taken with an FEI Quanta-600 FEG Environmental SEM using an acceleration voltage of 30 kV. EBSD measurements (Oxford Instruments) were obtained on this SEM system using an accelerating voltage of the primary electron beam for the backscatter images of 30 kV. Raman spectra (WITec Alpha 300 micro-Raman imaging system) were obtained using a 488-nm wavelength incident laser. AFM images were obtained with a PSIA model XE-100S using noncontact mode at 300 kHz.

Figure 1(a) shows a DF optical image of a graphene domain on Cu foil obtained immediately (within 10 min) after removal from the (cooled, 'H$_2$ protected') tube furnace when the growth was completed. The BF optical image acquired from the same area is shown in Fig. 1(b) for comparison. Both images were captured at the best condition of our optical microscopy system without any post-processing. The BF image shows a relatively uniform contrast, however, the DF image shows a 'flower-like' domain composed of bright ripples against a dark background. The 'flower-like' domain and the ripples match well to the dark region and the copper steps, respectively, in the SEM image captured from the same area [Fig. 1(c)]. The bright region in Fig. 1(a) is assigned to graphene, as is confirmed by the Raman mapping (intensity of G band) and Raman spectra shown in Fig. 1(d). The Raman spectrum [red curve in Fig. 1(d)] acquired in the region indicated by the red circle shows the typical G and 2D bands with no detectable D band, demonstrating high quality graphene. In contrast, the Raman spectrum [black dashed curve in Fig. 1(d)] acquired from the area indicated by the white dashed circle shows weak peaks centered at 154, 214, and 644 cm$^{-1}$ that are assigned to Cu$_2$O[13, 15], indicating a bare Cu area with onset of oxidation. We analyzed the imaging contrast of graphene on Cu foils by comparing the grayscale histograms of the DF, BF, and SEM images, which are shown in the lower panels of Fig. 1(a)-1(c), respectively. The grayscale histogram of the graphene (red) is very close to that of bare Cu (black) for the BF image, making graphene barely identifiable, but for the DF image the gray scale histogram of the graphene significantly differs from the bare Cu allowing ready identification of graphene. The full width at half maximum (FWHM) of the graphene grayscale histogram for the DF image is larger than that of the SEM image, which may be due to the ripples present in the graphene image.

Figure 2 shows the DF and BF images of freshly prepared graphene on Cu foils with different coverages, for identical regions. For the 90% covered graphene sample, graphene edges and boundaries between grains are distinguishable in the DF image as indicated by the red arrows [Fig. 2(a)], while it is difficult to identify graphene in the BF image [Fig. 2(b)] of the same region. For the fully-covered graphene on Cu foil [Fig. 2(c)], although there are no bare Cu areas providing the dark background to highlight graphene domains, the bright 'ripples' arising from the copper steps and the black lines (indicated by the red arrows) showing graphene wrinkles or cracks reveal the existence of graphene on top. In contrast, the DF image of bare Cu foil [supporting information, Fig. S1(a)] is completely different from that of fully-covered graphene on Cu [Fig. 2(c)], as it is dark except at the Cu grain boundaries and at some large dots that we assume are contaminants. The BF image of a fully-covered graphene sample [Fig. 2(d)] is very similar to that of bare Cu foil [Fig. S1(b)].

The contrast of graphene on Cu foils with BF imaging can be, as reported, enhanced by thermal annealing (Fig. S2).[5, 11] After heating the samples in air on a hot plate at 160 °C for 30 min, the BF image of graphene on Cu foil with 90% coverage [Fig. S2(b)] provides as high a contrast as the DF images [Fig. 2(a), Fig. S2(a)]. The brightness and color contrast with white light illumination arise from the light absorption that is dependent upon the refractive index and the opacity of the specimen.[16] Thermal annealing transformed the bare Cu regions to copper oxides and increased the interference color contrast between copper oxides and Cu, thus making graphene visible under bright field conditions. However, with this treatment it is still difficult to identify fully-covered graphene on Cu foils with BF imaging [Fig. S2(d)] as the graphene coating protects against oxide formation.[13] In any case heating graphene in air after growth could eventually damage it and since DF imaging works well, this is not needed for full coverage (Fig. S3).

Rather than illuminating the sample with a filled cone of light as in BF imaging, in DF imaging the condenser forms a hollow cone of light. The objective lens is located in the dark hollow of this cone, and the image is made only by those light rays scattered by the sample.[16] When a particle is small compared with the incident light wavelength, that is,

$$\alpha = \frac{\pi D}{\lambda} \leqslant 1 \tag{1}$$

where $\alpha$ is the ratio of the circumference of the particle ($\pi D$) to the wavelength of light ($\lambda$), the collected light is by Rayleigh scattering.[17] The scattered light intensity, $I$, is

$$I \propto \frac{D^6}{\lambda^4} \left|\frac{n^2-1}{n^2+2}\right| (1 + \cos^2 \theta) I_0 \tag{2}$$

where $I_0$ is the incident light intensity, $n$ is the refractive index of the particle, and $\theta$ is the scattering angle.[17] The intensity $I$ of the scattered light thus varies as the sixth order of the particle size $D$, that is DF is quite sensitive to relatively small variations in surface topology. According to our previous studies, copper steps that are formed during the graphene growth/cooling down process survive exposure to ambient due to the graphene coating, but there are no steps in bare Cu areas because of oxidation.[3, 13] In the DF image shown in Fig. 1(a), the whole graphene domain has bright ripples that well match the copper steps in the SEM image [Fig. 1(c)] of the same region. The height of the Cu steps beneath the graphene measured by atomic force microscopy (AFM) is 20±10 nm [red line in Fig. 3(a)]. Under visible light illumination with wavelength in the 390 to 700 nm range, the light is thus Rayleigh scattered by the copper steps. Compared to the bare Cu regions with its fairly low roughness of ~3 nm [black line in Fig. 3(a)], the light scattered by the ~20-nm high copper steps is roughly 88,000 times higher than that scattered by the bare Cu, and this is why high contrast is obtained with DF imaging.

In addition to the graphene domains having bright ripples, some graphene domains have bright edges superimposed on a dark background with little to no ripples under DF [such as graphene on the lower Cu grain in Fig. 3(b)]. These two types of DF imaging features were typically observed when there were two adjacent copper grains with different orientations (for this particular case, (110), upper; (138), lower) that were indexed by electron backscatter diffraction [EBSD, Fig. 3(b)]. The AFM image [Fig. 3(c)] captured at an 'edges highlighted' graphene domain had Cu step heights of around 3 nm (red line) thus comparable to the surface roughness of bare Cu (black line). However, the graphene edges had a relatively higher surface roughness of around 8 nm as measured by AFM [green line in Fig. 3(c)], which rationalizes the bright edges in DF imaging. High magnification DF images of individual graphene domains with the two different types of features and their corresponding SEM images are shown in Fig. 3(d). Cracks and sub-micron contaminants indicated by the arrows in Fig. 3(d) can be identified in both DF images (left) with a resolution comparable to the respective SEM images (right). For the graphene domain with 'edges highlighted' [lower left panel of Fig. 3(d)], the cracks and domain edges are brighter (i.e., 'highlighted') because of a higher surface roughness from oxidation of copper in these areas, as indicated by a Raman spectrum typical of those obtained at the edges of such graphene domains (Fig. S4).

We observed that graphene adlayers could be readily identified on Cu foils by DF imaging when the height of the copper steps varies rather dramatically with the thickness of the graphene film in that region. Figure 4(a) shows an SEM image of sub-monolayer graphene with adlayers (the darker regions) located at the center of each domain. Figure 4(b) shows a DF image obtained from the same region that has bright contours around the graphene domains (i.e., the edges of the domains) and bright regions with ripples matching the graphene adlayers. The DF image of an individual graphene domain is shown in Fig. 4(c). Copper steps are not visible in the region covered by single-layer graphene while they are observed beneath the graphene adlayer in the center. AFM [Fig. 1(d)] indicates that the copper steps beneath the central adlayer are around 12 nm in height (red line), higher than those beneath the surrounding single-layer graphene (less than 4 nm, green line), thus allowing the identification of graphene adlayers by DF imaging. We note that graphene adlayers of freshly prepared samples are only visible under DF in those domains where the single-layer graphene has 'edges highlighted'. For those domains that cover relatively large Cu steps such as with heights of about 20 nm, it was however hard to distinguish adlayers from single layers because of the very high contrast all over the graphene regions, as shown in Fig. 1(a) and 3(d) (upper left panel).

In conclusion, compared to the more commonly-used bright field imaging, optical microscopy with dark field imaging provides contamination-free, non-destructive, rapid, and facile evaluation of freshly prepared graphene directly on Cu foils without any additional sample treatment. Graphene on Cu foils with either partial or full coverage can be readily identified and the coverage level of graphene samples of several square centimeters can be obtained in minutes. The intensity of Rayleigh scattering by the copper steps beneath the graphene rationalizes the features of DF images of graphene. Cracks and grain boundaries as well as adlayers can be distinguished under DF due to variations in surface roughness. DF optical imaging can be readily extended to studies of graphene directly on other metallic growth substrates and is likely to be useful for other 2D materials such as we have also observed for thin films of h-BN on metal substrates (Fig. S5).

**Acknowledgments**

We appreciate support from the W. M. Keck Foundation, the Office of Naval Research, and the Army Research Office (Grant W911NF1010428), and comments by Dr. L. Colombo.

**Supporting Information Available**

Figure S1: DF and BF optical images of bare Cu foils.

Figure S2: The corresponding DF and BF optical images of Fig. 2 acquired after thermal annealing.

Figure S3: DF and BF optical images of graphene on Cu foils with partial coverage examined after heating in air for different time.

Figure S4: Raman spectrum typical of those detected at the edges of graphene domains on Cu foils.

Figure S5: DF and BF optical images of thin h-BN films on Ni foils.

**Figures and Captions**

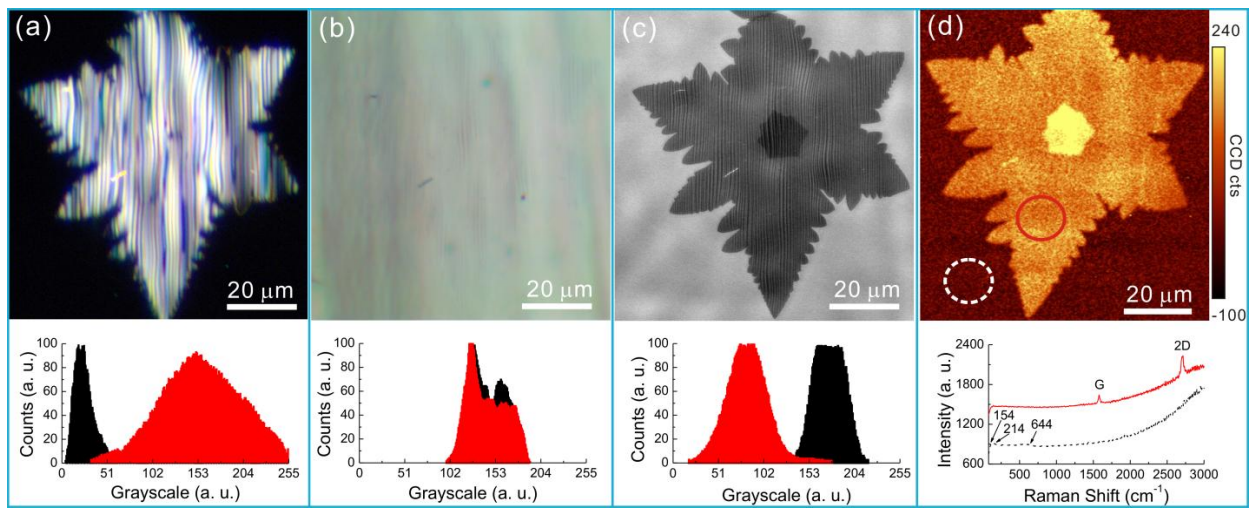

**FIG. 1.** An individual graphene domain on Cu foil (partial coverage) examined immediately after CVD growth. (a) Dark field (DF) optical image, (b) bright field (BF) optical image, (c) SEM image, and their corresponding grayscale histograms. Black columns represent bare Cu, and red graphene. (d) Raman map (intensity at 1580 cm$^{-1}$, G band) and spectra acquired in the graphene region circled in red (red curve) and in the bare copper area circled in dashed white (black dashed curve).

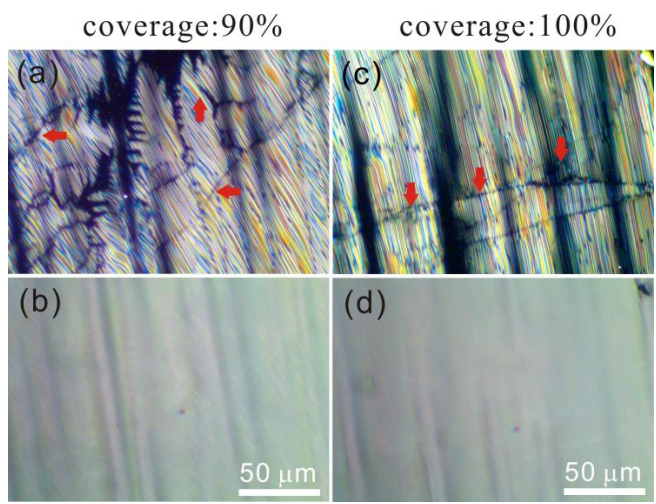

**FIG. 2.** (a, c) DF optical images of graphene on Cu foils with 90% (a) and 100% (c) coverages. Their corresponding BF images acquired from the same regions are shown in panels b and d, respectively.

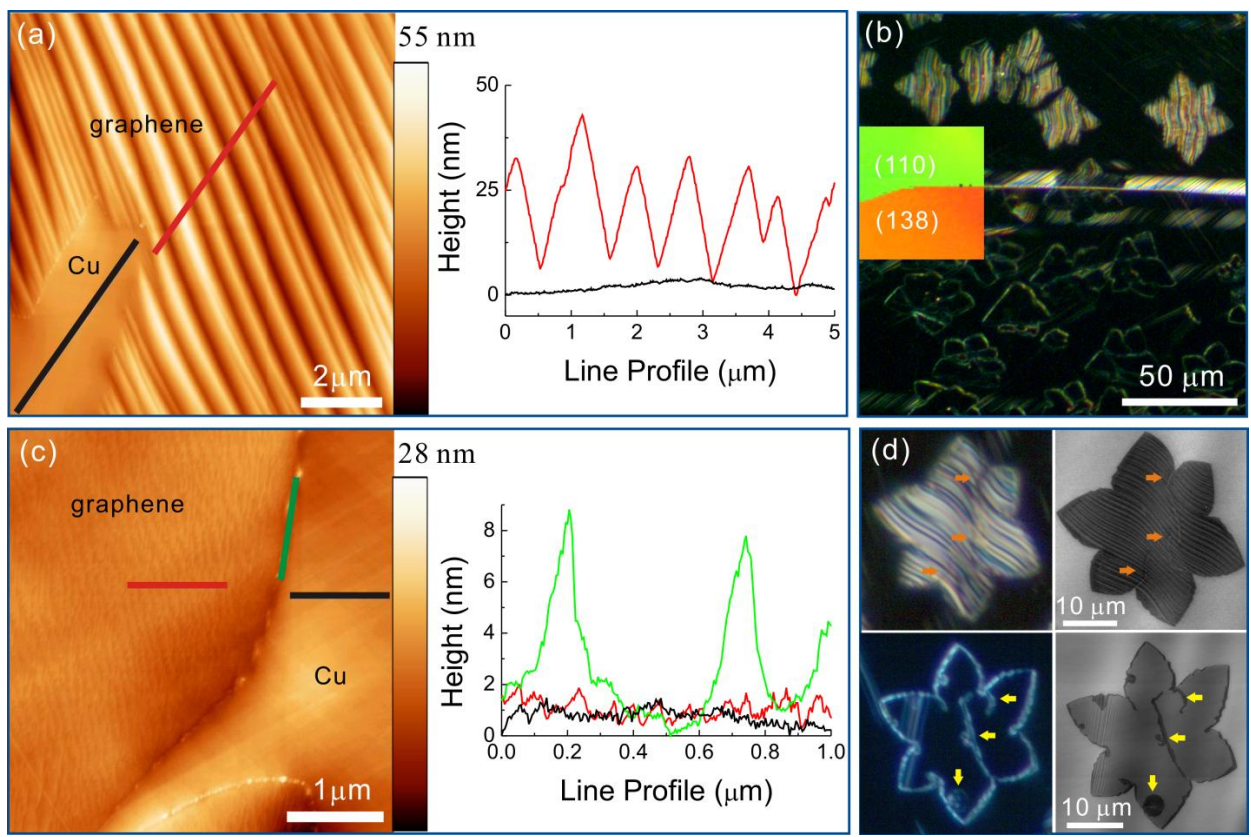

**FIG. 3.** (a, c) AFM topography images and their corresponding line profiles detected at 'Cu steps highlighted' and 'edges highlighted' graphene domains, respectively. (b) DF optical image of graphene domains with two different features present (as discussed in the text). The electron backscatter diffraction (EBSD) normal IPF image (insert) shows the orientations of the two copper grains. (d) High magnification DF images (left) of these two kinds of graphene domains and their corresponding SEM images (right).

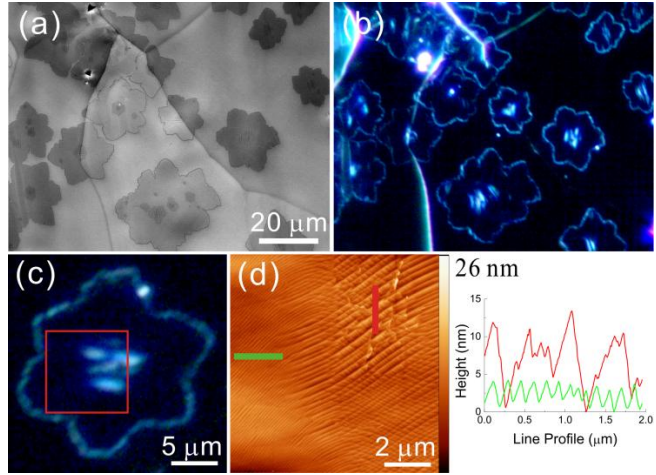

**FIG. 4.** (a) SEM and (b) DF images of graphene on Cu foil (partial coverage) with adlayers present at each graphene domain. (c) High magnification DF image of a graphene domain with adlayers at roughly the center. (d) AFM topography image and its line profiles acquired from the area within the red square in panel c.

**Supporting Information**

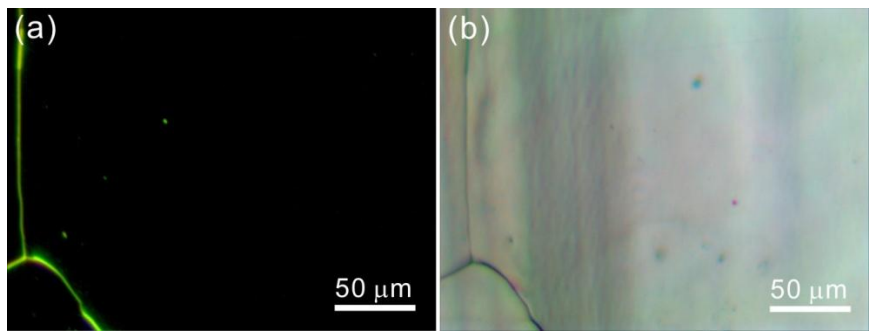

**FIG. S1.** Dark field (a) and its corresponding bright field (b) optical images of bare Cu foil without graphene coating. Only copper grain boundaries and several large dots (contamination) are visible in both images.

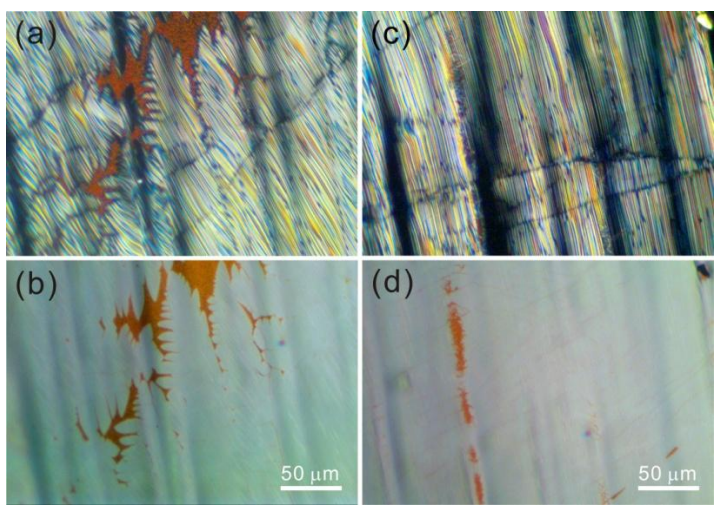

**FIG. S2.** DF (a, c) and their corresponding BF (b, d) optical images of graphene on Cu foils with 90% (a, b) and 100% (c, d) coverage, respectively, with images acquired after heating in air on a hot plate at 160 °C for 30 min. The images were captured for the same regions shown in Fig. 2. Sub-monolayer graphene domains are distinguishable by both DF and BF optical imaging after thermal annealing, while fully-coverage graphene can be identified only by DF imaging.

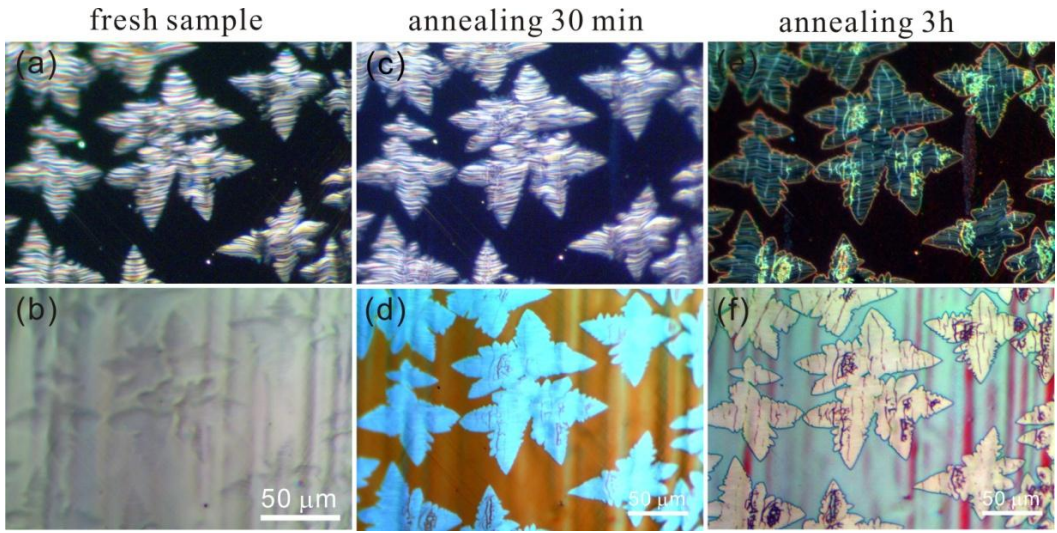

**FIG. S3.** DF (a, c, e) and their corresponding BF (b, d, f) optical images of graphene on Cu foils at partial coverage, examined after heating in air on a hot plate at 160 °C for the indicated times. More cracks in graphene domains are distinguished in both DF and BF optical images for the 3h vs the 30min annealing times. All the images were captured from the same region.

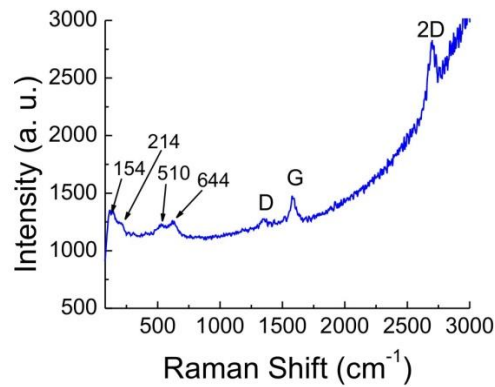

**FIG. S4.** Raman spectra acquired at the edges of a typical graphene domain on Cu foil. The Raman peaks (154, 214, 510, 644 cm$^{-1}$) are assigned to $Cu_2O$[15] and indicate the onset of oxidation at the graphene domain edges.

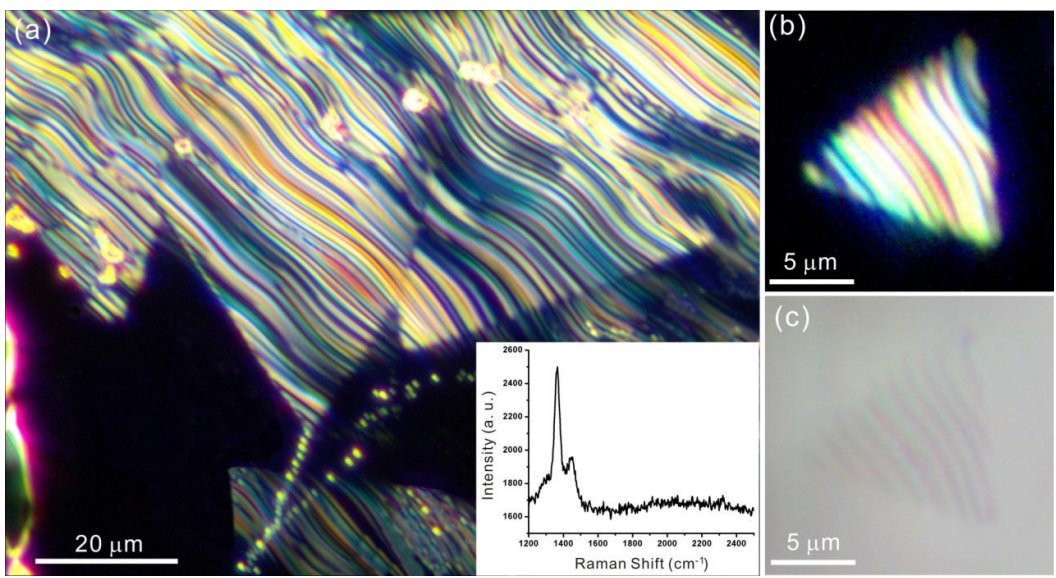

**FIG. S5.** (a) DF optical image of h-BN with partial coverage on a Ni foil (50-μm thick, 99.9% Goodfellow Corporation NI000340). The typical Raman spectrum of the h-BN films is shown in the insert. (b) DF and (c) BF optical images of a tiny h-BN triangle domain on Ni foil. (Details of the h-BN growth will be published soon and follows our previous report.) [S1]

**Supporting References**